\documentclass[a4paper,showpacs,showkeys,nofootinbib,aps]{revtex4}
\usepackage{epsfig}
\begin{document}
\boldmath
\title{The Ratio $R=F_L/F_T$ in DIS as a Probe\\
of the Charm Content of the Proton}
\unboldmath
\author{N.Ya.~Ivanov}
 \email{nikiv@mail.yerphi.am}
\affiliation{Yerevan Physics Institute, Alikhanian Br.~2, 375036 Yerevan, Armenia}

\begin{abstract}
\noindent We analyze the Callan-Gross ratio $R(x,Q^2)=F_L/F_T$ in heavy-quark leptoproduction as a 
probe of the charm content of the proton. To estimate the charm-initiated contributions, we use the 
ACOT($\chi$) variable-flavor-number scheme. Our analysis shows that charm densities of the recent CTEQ 
sets of parton distributions have sizeable impact on the Callan-Gross ratio in a wide region of $x$ 
and $Q^{2}$. In particular, the ACOT($\chi$) predictions for the quantity $R(x,Q^{2})$ are about half 
as large as the corresponding expectations of the photon-gluon fusion mechanism for $x\sim 10^{-2}-10^{-1}$
and $Q^2\gg m^2$. This is because the structure functions $F_{T}(x,Q^{2})$ and 
$F_{L}(x,Q^{2})$ have different dependences on the mass logarithms of the type $\alpha_{s}\ln\left( Q^{2}/m^{2}\right)$. On the other hand, our recent studies indicate that, contrary to the production 
cross sections, the Callan-Gross ratio is sufficiently stable under radiative corrections to the 
photon-gluon fusion component for $x\gtrsim 10^{-4}$. We conclude that the quantity $R(x,Q^{2})$ in 
heavy-quark leptoproduction is perturbatively stable but sensitive to resummation of the mass logarithms. 
For this reason, in contrast to the structure functions, the ratio $R(x,Q^{2})=F_L/F_T$  could be good 
probe of the charm density in the proton.
\end{abstract}
 \maketitle

\section{Introduction}
The notion of the intrinsic charm (IC) content of the proton has been introduced over 25 years ago
in Ref.~\cite{BHPS}. It was shown that, in the light-cone Fock space picture
\cite{brod1}, it is natural to expect a five-quark state contribution, $\left\vert
uudc\bar{c}\right\rangle$, to the proton wave function. This component can be generated by
$gg\rightarrow c\bar{c}$ fluctuations inside the proton where the gluons are coupled to different
valence quarks. The original concept of the charm density in the proton \cite{BHPS} has
nonperturbative nature since a five-quark contribution $\left\vert uudc\bar{c}\right\rangle$ scales
as $1/m^{2}$ where $m$ is the $c$-quark mass \cite{polyakov}.

In the middle of nineties, another point of view on the charm content of the proton has been proposed 
in the framework of the variable-flavor-number scheme (VFNS) \cite{ACOT,collins}.  The VFNS is an 
approach alternative to the traditional fixed-flavor-number scheme (FFNS) where only light degrees of
freedom ($u,d,s$ and $g$) are considered as active. Within the VFNS, the
mass logarithms of the type $\alpha_{s}\ln\left( Q^{2}/m^{2}\right)$ are resummed through the all
orders into a heavy quark density which evolves with $Q^{2}$ according to the standard DGLAP \cite{grib-lip}
evolution equation. Hence this approach introduces the parton distribution functions (PDFs) for the
heavy quarks and changes the number of active flavors by one unit when a heavy quark threshold is
crossed. Note also that the charm density arises within the VFNS perturbatively via the
$g\rightarrow c\bar{c}$ evolution. Some recent developments concerning the VFNS are presented in
Refs.~\cite{chi,SACOT,Thorne-NNLO}.

Presently, both nonperturbative IC and perturbative charm density are widely used for a
phenomenological description of available data. (A recent review of the theory and experimental
constraints on the charm quark distribution may be found in Ref.~\cite{pumplin}). In particular, 
practically all the recent versions of the CTEQ \cite{CTEQ4,CTEQ5,CTEQ6} and MRST \cite{MRST2004}
sets of PDFs are based on the VFN schemes and contain a charm density. At the same time, the key
question remains open: How to measure the charm content of the proton? The basic theoretical
problem is that radiative corrections to the heavy-flavor production cross sections are large: 
they increase the leading order (LO) results by approximately a factor of two. Moreover, 
soft-gluon resummation of the threshold Sudakov logarithms indicates
that higher-order contributions can also be substantial. (For reviews, see 
Refs.~\cite{Laenen-Moch,kid1}.) On the other hand, perturbative instability leads to a high 
sensitivity of the theoretical calculations to standard uncertainties in the input QCD parameters: 
the heavy-quark mass, $m$, the factorization and renormalization scales, $\mu _{F}$ and $\mu _{R}$, 
the asymptotic scale parameter $\Lambda_{\mathrm{QCD}}$ and the PDFs. For this reason, one can
only estimate the order of magnitude of the pQCD predictions for charm production cross sections in 
the entire energy range from the fixed-target experiments \cite{Mangano-N-R} to the RHIC collider 
\cite{R-Vogt}.

Since production cross sections are not perturbatively stable, they cannot be a good probe of the
charm density in the proton.\footnote{It will be shown bellow that, in a wide kinematic range, the 
heavy-flavor-initiated contributions have approximately the same effect on the structure function 
$F_{2}(x,Q^{2})$ as the radiative corrections to the dominant photon-gluon fusion mechanism.} 
For this reason, it is of special interest 
to study those observables that are well-defined in pQCD. Nontrivial examples of such observables were 
proposed in Refs.~\cite{we1,we2,we3,we4,we5,we7}, where the azimuthal $\cos(2\varphi)$ asymmetry and 
Callan-Gross ratio $R(x,Q^2)=F_L/F_T$  in heavy quark leptoproduction were analyzed.\footnote{
Well-known examples include the shapes of differential cross sections
of heavy flavor production, which are sufficiently stable under radiative
corrections.}$^{,}$\footnote{Note also the recent paper \cite{Almeida-S-V}, where the
perturbative stability of the QCD predictions for the charge asymmetry in top-quark
hadroproduction has been observed.} In particular, the Born-level results were
considered \cite{we1} and the NLO soft-gluon corrections to the basic mechanism,
photon-gluon fusion (GF), were calculated \cite{we2,we4}. It was shown that, contrary to
the production cross sections, the azimuthal asymmetry in heavy flavor photo- and
leptoproduction is quantitatively well defined in pQCD: the contribution of the dominant
GF mechanism to the asymmetry is stable, both parametrically and
perturbatively. Therefore, measurements of this asymmetry should provide a clean test of
pQCD. As was shown in Ref.~\cite{we3}, the azimuthal asymmetry in open charm
photoproduction could be measured with an accuracy of about ten percent in the
approved E160/E161 experiments at SLAC \cite{E161} using the inclusive spectra of
secondary (decay) leptons.

In Ref.~\cite{we5}, the photon-(heavy) quark scattering (QS) contribution to 
$\varphi$-dependent lepton-hadron deep-inelastic scattering (DIS) was investigated. It
turned out that, contrary to the basic photon-gluon fusion component, the QS
mechanism is practically $\cos(2\varphi)$-independent. This is due to the fact that the
quark-scattering contribution to the $\cos(2\varphi)$ asymmetry is, for
kinematic reasons, absent at LO and is negligibly small at NLO, of the order of $1\%$.
This indicates that the azimuthal distributions in charm leptoproduction could be good
probe of the charm PDF in the proton.

The perturbative and parametric stability of the GF predictions for the Callan-Gross ratio 
$R(x,Q^2)=F_L/F_T$ in heavy-quark leptoproduction was considered in Ref.~\cite{we7}. It was 
shown that large radiative corrections to the structure functions $F_T(x,Q^2)$ and 
$F_L(x,Q^2)$ cancel each other in their ratio $R(x,Q^2)$ with good accuracy. As a result, 
the next-to-leading order (NLO) contributions of the dominant GF mechanism to the Callan-Gross 
ratio are less than $10\%$ in a wide region of the variables $x$ and $Q^2$.

In the present paper, we continue the studies of the heavy-quark-initiated contributions to
heavy-flavor production in DIS:
\begin{equation}
\ell(l )+N(p)\rightarrow \ell(l -q)+Q(p_{Q})+X[\bar{Q}](p_{X}). \label{1}
\end{equation}
In the case of unpolarized initial states and neglecting the contribution of $Z$-boson exchange,
the cross section of reaction (\ref{1}) can be written as
\begin{eqnarray}
\frac{\text{d}^{2}\sigma_{lN}}{\text{d}x\,\text{d}Q^{2}}&=&\frac{4\pi
\alpha^{2}_{\mathrm{em}}}{Q^{4}}\left\{ \left[ 1+(1-y)^{2}\right] F_{T}( x,Q^{2})
+2\left(1-y\right) F_{L}(x,Q^{2})\right\}  \nonumber \\
&=&\frac{2\pi \alpha^{2}_{\mathrm{em}}}{xQ^{4}}\left\{ \left[ 1+(1-y)^{2}\right] F_{2}( x,Q^{2})
-2xy^{2} F_{L}(x,Q^{2})\right\},  \label{2}
\end{eqnarray}
where $\alpha_{\mathrm{em}}$ is Sommerfeld's fine-structure constant,
$F_{2}(x,Q^2)=2x(F_{T}+F_{L})$ and the kinematic variables are defined by
\begin{eqnarray}
\bar{S}=\left( \ell +p\right) ^{2},\qquad &Q^{2}=-q^{2},\qquad &x=\frac{Q^{2}}
{2p\cdot q},  \nonumber \\
y=\frac{p\cdot q}{p\cdot \ell },\qquad \quad ~ &Q^{2}=xy\bar{S},\qquad &\xi=
\frac{Q^2}{m^2}.  \label{3}
\end{eqnarray}
In this paper, we investigate the QS contribution to the Callan-Gross ratio
in heavy-quark leptoproduction defined as
\begin{equation}
R(x,Q^{2})=\frac{F_{L}(x,Q^{2})}{F_{T}(x,Q^{2})}. \label{4}
\end{equation}
To estimate the charm-initiated contributions to the ratio $R(x,Q^{2})$, we use the ACOT($\chi$) 
VFNS proposed in Ref.~\cite{chi}. Our analysis shows that charm densities of the recent 
CTEQ \cite{CTEQ4,CTEQ5,CTEQ6} sets of PDFs lead to a sizeable reduction of the 
GF predictions for the Callan-Gross ratio at $x>10^{-4}$. 
For instance, the ACOT($\chi$) VFNS predictions for the ratio $R(x,Q^{2})$ are about half of 
the corresponding FFNS ones for $x\sim 10^{-2}$--$10^{-1}$ and $Q^2\gg m^2$.
This is due to the fact that resummation of the mass logarithms has different effects on
the structure functions $F_{T}(x,Q^{2})$ and $F_{L}(x,Q^{2})$ because they have different 
dependences on the  quantities $\alpha_s^n\ln^k (Q^{2}/m^{2})$. In particular, contrary to 
the transverse structure function, $F_{T}(x,Q^{2})$, the longitudinal one, $F_{L}(x,Q^{2})$, 
does not contain potentially large mass logarithms at both LO and NLO \cite{LRSN,BMSMN}.

On the other hand, our recent studies indicate that radiative corrections to the Callan-Gross 
ratio do not exceed $10\%$ for $x\gtrsim 10^{-4}$ practically at all values of $Q^2$ \cite{we7}. 
We conclude that the quantity $R(x,Q^{2})$ in 
heavy-quark leptoproduction is perturbatively stable but sensitive to resummation of the mass 
logarithms of the type $\alpha_s\ln (Q^{2}/m^{2})$. For this reason, in contrast to the structure 
functions, the ratio $R(x,Q^{2})=F_L/F_T$ in DIS could be good probe of the charm density in the 
proton.

Concerning the experimental aspects, the ratio $R(x,Q^{2})$ in charm leptoproduction can, in
principle, be measured in future studies at the proposed eRHIC \cite{eRHIC} and 
LHeC \cite{LHeC} colliders at BNL and CERN, correspondingly.

This paper is organized as follows. In Section~\ref{parton}, we briefly discuss the GF and QS 
predictions for the parton-level cross sections. Resummation of the mass logarithms for the transverse
and longitudinal structure functions within the ACOT($\chi$) VFNS is considered in Section~\ref{resum}. 
Hadron-level predictions of both FFNS and VFNS for the structure function $F_{2}(x,Q^{2})$ and Callan-Gross 
ratio $R(x,Q^2)=F_L/F_T$ in charm leptoproduction are discussed in Section~\ref{hadron}.

\section{\label{parton} Parton-Level Cross Sections}

\subsection{Born-Level Results}
\begin{figure}
\begin{center}
\mbox{\epsfig{file=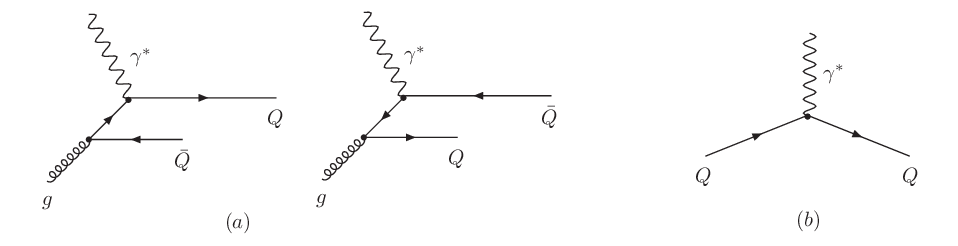,width=445pt}}
\end{center}
\caption{\label{Fg.1}\small LO Feynman diagrams of the photon-gluon fusion (\emph{a}) 
and photon-quark scattering (\emph{b}).}
\end{figure}
At LO, ${\cal O}(\alpha_{\mathrm{em}}\alpha_{s})$, the photon-gluon component of the heavy-quark 
leptoproduction is described by the following parton-level interaction:
\begin{equation} \label{5}
\gamma ^{*}(q)+g(k_{g})\rightarrow Q(p_{Q})+\bar{Q}(p_{\bar{Q}}).
\end{equation}
The relevant Feynman diagrams are depicted in Fig.~\ref{Fg.1}\emph{a}.
The LO $\gamma ^{*}g$ cross sections, $\hat{\sigma}_{k,\mathrm{g}}^{(0)}(z,\lambda)$ ($k=2,L$), 
have the form \cite{LW1}:
\begin{eqnarray}
\hat{\sigma}_{2,\mathrm{g}}^{(0)}(z,\lambda)&=&\frac{\alpha_{s}(\mu_R^2)}{2\pi}\hat{\sigma}_{B}(z)
\left\{\left[(1-z)^{2}+z^{2}+4\lambda z(1-3z)-8\lambda^{2}z^{2}\right]
\ln\frac{1+\beta_{z}}{1-\beta_{z}}-
\left[1+4z(1-z)(\lambda-2)\right]\beta_{z}\right\},  \label{6}\\
\hat{\sigma}_{L,\mathrm{g}}^{(0)}(z,\lambda)&=&\frac{2\alpha_{s}(\mu_R^2)}{\pi}\hat{\sigma}_{B}(z)z
\left\{-2\lambda z\ln\frac{1+\beta_{z}}{1-\beta_{z}}+\left(1-z\right)\beta_{z}\right\},
\label{7}
\end{eqnarray}
with
\begin{equation}\label{8}
\hat{\sigma}_{B}(z)=\frac{(2\pi)^2e_{Q}^{2}\alpha_{\mathrm{em}}}{Q^{2}}z,
\end{equation}
where $e_{Q}$ is the electric charge of quark $Q$ in units of the positron charge and
$\alpha_{s}(\mu_R^2)$ is the strong-coupling constant.
In Eqs.~(\ref{6})--(\ref{8}), we use the following definition of partonic kinematic
variables:
\begin{equation}\label{9}
z=\frac{Q^{2}}{2q\cdot k_{g}},\qquad\lambda =\frac{m^{2}}{Q^{2}}, \qquad
\beta_{z}=\sqrt{1-\frac{4\lambda z}{1-z}}.
\end{equation}
Within the FFNS, corresponding hadron-level cross sections, $\sigma_{k,\mathrm{GF}}(x,Q^2)$ ($k=2,T,L$), 
have the form
\begin{equation}
\sigma_{k,\mathrm{GF}}(x,Q^2)=\int\limits_{x(1+4\lambda)}^{1}\text{d}z\,g(z,\mu_{F})
\hat{\sigma}_{k,\mathrm{g}}\left(\frac{x}{z},\lambda,\mu_{F},\mu_{R}\right),
\label{10}
\end{equation}
where $g(z,\mu_{F})$ is the gluon PDF of the proton. The leptoproduction cross sections
$\sigma_{k}(x,Q^2)$ are related to the structure functions $F_{k}(x,Q^2)$ 
as follows:
\begin{eqnarray}
F_{k}(x,Q^2) &=&\frac{Q^{2}}{8\pi^{2}\alpha_{\mathrm{em}}x}\sigma_{k}(x,Q^2)
\qquad (k=T,L), \label{11}\\
F_{2}(x,Q^2) &=&\frac{Q^{2}}{4\pi^{2}\alpha_{\mathrm{em}}}\sigma_{2}(x,Q^2),
\label{11a}
\end{eqnarray}
where 
\begin{equation}\label{11b}
\sigma_{T}(x,Q^2)=\sigma_{2}(x,Q^2)-\sigma_{L}(x,Q^2).
\end{equation}

At leading order, ${\cal O}(\alpha _{em})$, the only quark scattering subprocess is 
\begin{equation}
\gamma ^{*}(q)+Q(k_{Q})\rightarrow Q(p_{Q}).  \label{12}
\end{equation}
Corresponding Feynman diagram is depicted in Fig.~\ref{Fg.1}\emph{b}.
The LO $\gamma ^{*}Q$ cross sections, $\hat{\sigma}_{k,\mathrm{Q}}^{(0)}(z,\lambda)$ ($k=2,L$), 
are \cite{we5}:
\begin{eqnarray}
\hat{\sigma}_{2,\mathrm{Q}}^{(0)}(z,\lambda)&=&\hat{\sigma}_{B}(z)\sqrt{1+4\lambda z^{2}}\,
\delta(1-z), \label{13}\\
\hat{\sigma}_{L,\mathrm{Q}}^{(0)}(z,\lambda)&=&\hat{\sigma}_{B}(z)\frac{4\lambda z^{2}}
{\sqrt{1+4\lambda z^{2}}}\,\delta(1-z), \label{14}
\end{eqnarray}
with $z=Q^{2}/(2q\cdot k_{Q})$.

\subsection{NLO Corrections}

At NLO, ${\cal O}(\alpha_{\mathrm{em}}\alpha_{s}^2)$, the contribution of the photon-gluon
component is usually presented in terms of the dimensionless coefficient functions
$c_{k,\mathrm{g}}^{(n,l)}(z,\lambda)$ ($k=T,L$), as
\begin{equation}\label{15}
\hat{\sigma}_{k,\mathrm{g}}(z,\lambda,m^2,\mu^{2})=\frac{e_{Q}^{2}\alpha_{\mathrm{em}}\alpha_{s}(\mu
^{2})}{m^{2}}\left\{c_{k,\mathrm{g}}^{(0,0)}(z,\lambda)+ 4\pi\alpha_{s}(\mu^{2})\left[
c_{k,\mathrm{g}}^{(1,0)}(z,\lambda)+c_{k,\mathrm{g}}^{(1,1)}(z,\lambda)\ln\frac{\mu^{2}}{m^{2}}
\right]\right\}+{\cal O}(\alpha_{s}^2).
\end{equation}
where we identify $\mu=\mu_{F}=\mu_{R}$.

In this paper, we neglect the $\gamma ^{*}q(\bar{q})$ fusion subprocesses. This is
justified as their contributions to heavy-quark leptoproduction vanish at LO and are
small at NLO \cite{LRSN}. To be precise, the light-quark-initiated corrections to both
$F_T$ and $F_L$ structure functions are negative and less than $10\%$ in a wide
kinematic range \cite{LRSN}. Our estimates show that these contributions cancel in the 
ratio $R(x,Q^2)=F_L/F_T$ with an accuracy less than few percent. We also neglect the 
NLO corrections to the QS component due to their numerical insignificance \cite{HM,we5}.

The coefficients $c_{T,\mathrm{g}}^{(1,1)}(z,\lambda)$ and $c_{L,\mathrm{g}}^{(1,1)}(z,\lambda)$ 
of the $\mu$-dependent logarithms can be evaluated explicitly using renormalization group
arguments \cite{LRSN,Laenen-Moch}. The results of direct calculations of the coefficient
functions $c_{k,\mathrm{g}}^{(1,0)}(z,\lambda)$ ($k=T,L$) are presented in Refs.~\cite{LRSN,RSN}.

The analytic form of the heavy-quark coefficient functions for
lepton-hadron DIS in the kinematical regime $Q^2\gg m^2$ is presented in
Ref.~\cite{BMSMN}. The calculations were performed up to NLO
in $\alpha_s$ using operator product expansion techniques.\footnote{For the longitudinal 
cross section $\hat{\sigma}_{L,\mathrm{g}}(z,Q^2, m^2,\mu^{2})$, the asymptotic heavy-quark 
coefficient functions, $a_{L,\mathrm{g}}^{l,(n,m)}(z)$, are known up to NNLO in $\alpha_s$ 
\cite{Blumlein}.}
In the asymptotic regime $\xi \to \infty$, the production cross sections have the
following decomposition in terms of the coefficient functions $a_{k,\mathrm{g}}^{l,(n,m)}(z)$
($k=2,L$):
\begin{equation}\label{16}
\hat{\sigma}_{k,\mathrm{g}}(z,Q^2, m^2,\mu^{2})=\frac{e_{Q}^{2}\alpha_{\mathrm{em}}}{4\pi
m^{2}}\sum_{l=1}^{\infty }\left[ 4\pi \alpha_{s}(\mu ^{2})\right]
^{l}\sum_{m+n<l}^{n}a_{k,\mathrm{g}}^{l,(n,m)}(z)\ln^{n}\frac{\mu
^{2}}{m^{2}}\ln^{m}\frac{Q^{2}}{m^{2}}+{\cal O}
\left(\frac{m^{2}}{Q^{2}}\right).
\end{equation}

It was found in Refs.~\cite{BMSMN,BMSN} that the hadron-level structure function
$F^{\text{asymp}}_{2}(x,Q^2)$ approaches, to within ten percent, the corresponding exact
value $F^{\text{exact}}_{2}(x,Q^2)$ for $\xi \gtrsim 10$ and $x<10^{-1}$ both at LO and NLO.
In the case of the longitudinal structure function $F^{\text{asymp}}_{L}(x,Q^2)$, the
approach to $F^{\text{exact}}_{L}(x,Q^2)$ starts at much larger values of $\xi \gtrsim
4\times 10^{2}$.

\section{\label{resum} Resummation of the Mass Logarithms}

One can see from Eq.~(\ref{6}) that the GF cross section $\hat{\sigma}_{2,g}^{(0)}(z,\lambda)$
contains potentially large logarithm, $\ln (Q^{2}/m^{2})$. The same situation takes also place for
the NLO cross section $\hat{\sigma}_{2,g}^{(1)}(z,\lambda)$ \cite{LRSN,BMSMN}. At high
energies, $Q^{2}\rightarrow \infty$, the terms of the form $\alpha_{s}\ln (Q^{2}/m^{2})$ dominate
the production cross sections. To improve convergence of the perturbative series at high
energies, the so-called variable flavor number schemes (VFNS) have been proposed. Originally, this
approach was formulated in Refs.~\cite{ACOT,collins}.

In the VFNS, mass logarithms of the type $\alpha_s^n\ln^n (Q^{2}/m^{2})$ are resummed via the
renormalization group equations. In practice, the resummation procedure consists of two steps.
First, the mass logarithms have to be subtracted from the fixed order predictions for the partonic
cross sections in such a way that, in the asymptotic limit $Q^{2}\rightarrow \infty$, the well known
massless $\overline{\text{MS}}$ coefficient functions are recovered. Instead, a charm parton
density in the hadron, $c(x,Q^{2})$, has to be introduced. This density obeys the usual massless
NLO DGLAP \cite{grib-lip} evolution equation with the boundary condition $c(x,Q^{2}=Q_{0}^2)=0$ 
where $Q_{0}^2\sim m^{2}$. So, we may say that, within the VFNS, the charm density arises 
perturbatively from the $g\rightarrow c\bar{c}$ evolution.

In the VFNS, the treatment of the charm depends on the values chosen for $Q^{2}$. At low
$Q^{2}<Q_{0}^2$, the production cross sections are described by the light parton contributions
($u,d,s$ and $g$). The charm production is dominated by the GF process and its higher order QCD
corrections. At high $Q^{2}\gg m^{2}$, the charm is treated in the same way as the other light
quarks and it is represented by a charm parton density in the hadron, which evolves in $Q^{2}$. In
the intermediate scale region, $Q^{2}\sim m^{2}$, one has to make a smooth connection between the
two different prescriptions.

Strictly speaking, the perturbative charm density is well defined at high $Q^2\gg m^2$ but does not
have a clean interpretation at low $Q^2$. Since the charm distribution originates from resummation of
the mass logarithms of the type $\alpha_s^n\ln^n (Q^{2}/m^{2})$, it is usually assumed that the
corresponding PDF vanishes with these logarithms, i.e. for $Q^{2}<Q_{0}^2\approx m^{2}$. On the
other hand, the threshold constraint $W^2=(q+p)^2=Q^2(1/x-1)>4m^2$ implies that $Q_0$ is not a
constant but "live" function of $x$. To avoid this problem, several solutions have been proposed
(see e.g. Refs.~\cite{chi,SACOT}). In this paper, we use the so-called ACOT($\chi$) prescription
\cite{chi} which guarantees (at least at $Q^2>m^2$) the correct threshold behavior of the
heavy-quark-initiated contributions.

Within the VFNS, the charm production cross section has three pieces:
\begin{equation}\label{17}
\sigma_{2}(x,\lambda)=\sigma_{2,\mathrm{GF}}(x,\lambda)-\sigma_{2,\mathrm{SUB}}(x,\lambda)+
\sigma_{2,\mathrm{QS}}(x,\lambda),
\end{equation}
where the first and third terms on the right-hand side describe the usual (unsubtracted) GF and QS
contributions while the second (subtraction) term renders the total result infra-red safe in the
limit $m\rightarrow 0$. The only constraint imposed on the subtraction term is to reproduce at
high energies the familiar $\overline{\text{MS}}$ partonic cross section:
\begin{equation}\label{18}
\lim_{\lambda\rightarrow 0}\left[\hat{\sigma}_{2,\mathrm{g}}(z,\lambda)-
\hat{\sigma}_{2,\mathrm{SUB}}(z,\lambda)\right]=\hat{\sigma}^{\overline{\mathrm{MS}}}_{2,\mathrm{g}}(z).
\end{equation}
Evidently, there is some freedom in the choice of finite mass terms of the form $\lambda^{n}$ (with
a positive $n$) in $\hat{\sigma}_{2,\mathrm{SUB}}(z,\lambda)$. For this reason, several prescriptions 
have been proposed to fix the subtraction term. As mentioned above, we use the so-called ACOT($\chi$)
scheme \cite{chi}.

According to the ACOT($\chi$) prescription, the lowest order cross section is
\begin{eqnarray}
\sigma^{(\mathrm{LO})}_{2}(x,\lambda)&=&\int\limits_{\chi}^{1}\text{d}z\,g(z,\mu_{F})
\left[\hat{\sigma}_{2,\mathrm{g}}^{(0)}
\!\left(x/z,\lambda\right)-\frac{\alpha_{s}}{\pi}\ln\frac{\mu_{F}^{2}}{m^{2}}
\;\hat{\sigma}_{B}\left(x/z\right)P^{(0)}_{g\rightarrow c}\left(\chi/z\right)\right]+\hat{\sigma}_{B}(x)c_{+}(\chi,\mu_{F}), \label{19} \\ 
\chi&=&x(1+4\lambda), \label{19a}
\end{eqnarray}
where $P^{(0)}_{g\rightarrow c}$ is the LO gluon-quark splitting function, $P^{(0)}_{g\rightarrow
c}(\zeta)=\left.\left[(1-\zeta)^{2}+\zeta^{2}\right]\right/2$, $c_{+}(\zeta,\mu_{F})=c(\zeta,\mu_{F})+\bar{c}(\zeta,\mu_{F})$, and the LO GF cross section
$\hat{\sigma}_{2,\mathrm{g}}^{(0)}$ is given by Eq.~(\ref{6}). 

The asymptotic behavior of the subtraction terms is fixed by the parton level factorization
theorem. This theorem implies that the partonic cross sections d$\hat{\sigma}$ can be factorized
into process-dependent infra-red safe hard scattering cross sections d$\tilde{\sigma}$, which are
finite in the limit $m\rightarrow 0$, and universal (process-independent) partonic PDFs
$f_{a\rightarrow i}$ and fragmentation functions $d_{n\rightarrow Q}$:
\begin{equation}\label{add1}
\text{d}\hat{\sigma}(\gamma^{*}+a\rightarrow Q+X)=\sum_{i,n}f_{a\rightarrow
i}(\zeta)\otimes\text{d}\tilde{\sigma}(\gamma^{*}+i\rightarrow n+X)\otimes d_{n\rightarrow Q}(z).
\end{equation}
In Eq.~(\ref{add1}), the symbol $\otimes$ denotes the usual convolution integral, the indices
$a,i,n$ and $Q$ denote partons, $p_{i}=\zeta p_{a}$ and $p_{Q}=z p_{n}$. All the logarithms of the
heavy-quark mass (i.e., the singularities in the limit $m\rightarrow 0$) are contained in the PDFs
$f_{a\rightarrow i}$ and fragmentation functions $d_{n\rightarrow Q}$ while d$\tilde{\sigma}$ are
infra-safe (i.e., are free of the $\ln m^{2}$ terms). The expansion of Eq.~(\ref{add1}) can be used 
to determine order by order the subtraction terms. In particular, for the LO GF contribution to the
charm leptoproduction one finds \cite{ACOT}
\begin{equation}\label{add2}
\hat{\sigma}^{(0)}_{k,\mathrm{SUB}}\left(z,\ln\,(\mu_{F}^{2}/m^{2})\right)=f^{(1)}_{g\rightarrow
c}\left(\zeta,\ln\,(\mu_{F}^{2}/m^{2})\right)\otimes\hat{\sigma}^{(0)}_{k,\mathrm{Q}}(z/\zeta), 
\qquad \qquad  (k=2,L),
\end{equation}
where the quantity $f^{(1)}_{g\rightarrow c}\left(\zeta,\ln\,(\mu_{F}^{2}/m^{2})\right)=\left(
\alpha_{s}/2\pi\right)\ln\,(\mu_{F}^{2}/m^{2})\,P^{(0)}_{g\rightarrow c}\left(\zeta\right)$
describes the charm distribution in the gluon within the $\overline{\text{MS}}$ factorization
scheme.

One can see from Eq.~(\ref{add2}) that the longitudinal GF cross section
$\sigma_{L,\mathrm{GF}}(x,\lambda)$ does not have subtraction term at LO because the lowest 
order QS contribution $\hat{\sigma}_{L,\mathrm{Q}}^{(0)}(z,\lambda)$ given by Eq.~(\ref{14}) 
vanishes for $\lambda \to 0$. This is in accordance with infra-red behavior of the LO GF cross 
section $\hat{\sigma}_{L,\mathrm{g}}^{(0)}(z,\lambda)$ given by Eq.~(\ref{7}) which does not 
contain potentially large logarithms of the type $\ln (Q^{2}/m^{2})$.\footnote{Note that the 
NLO longitudinal cross section $\hat{\sigma}_{L,\mathrm{g}}^{(1)}(z,\lambda)$ is also 
infra-red safe \cite{LRSN,BMSMN}. Mass logarithms of the type $\ln (Q^{2}/m^{2})$
appear in the longitudinal GF structure function only at NNLO \cite{Blumlein}.} 
For this reason, the LO longitudinal cross section within the VFNS has the same form as 
in the FFNS:
\begin{equation}\label{20}
\sigma^{(\mathrm{LO})}_{L}(x,\lambda)=\int\limits_{\chi}^{1}\text{d}z\,g(z,\mu_{F})\,
\hat{\sigma}_{L,\mathrm{g}}^{(0)}\!\left(x/z,\lambda\right).
\end{equation}

In principle, one can add to the right-hand side of Eq.~(\ref{20}) mass terms of the form 
$\lambda^{n}$ (with a positive $n$), as was done in Refs.~\cite{chuvakin,Thorne-FL}. However, 
these terms are irrelevant for sufficiently high $Q^2$ where the VFNS is expected to be 
adequate.

\section{\label{hadron} Hadron-Level Predictions}

In this section, we present numerical analysis of the NLO corrections and charm-initiated 
contributions to the structure function $F_{2}(x,Q^{2})$ and Callan-Gross ratio $R(x,Q^2)=F_L/F_T$ 
in charm leptoproduction. In our calculations, we use the CTEQ5M parametrization of the gluon 
and charm PDFs together with the value $m_c=1.3$~GeV \cite{CTEQ5}.\footnote{Note that we convolve 
the NLO CTEQ5M gluon distribution function with both the LO and NLO partonic cross sections 
that makes it possible to estimate directly the degree of stability of the GF predictions under 
radiative corrections.}
The default value of the factorization and renormalization scales is $\mu=\sqrt{4m_{c}^{2}+Q^{2}}$.
\begin{figure}
\begin{center}
\begin{tabular}{cc}
\mbox{\epsfig{file=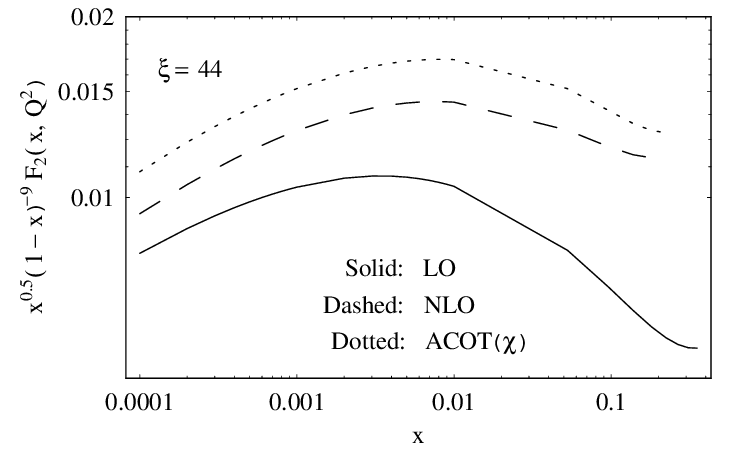,width=250pt}}
& \mbox{\epsfig{file=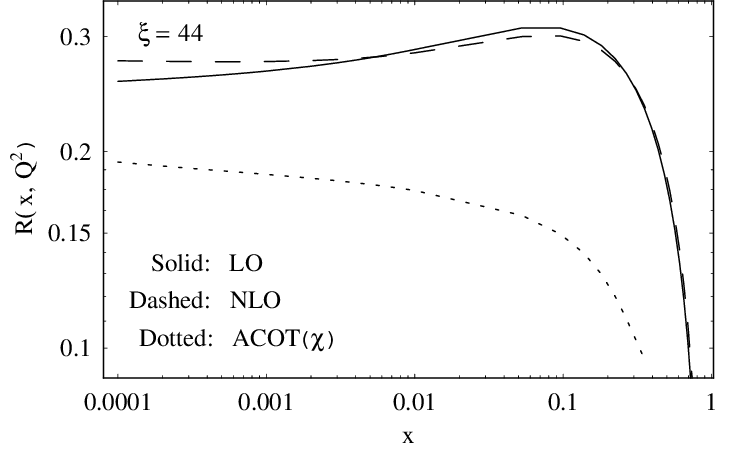,width=250pt}}\\
\end{tabular}
\caption{\label{Fg.2}\small \emph{Left panel:} $x$ dependence of the structure 
function $F_2(x,Q^2)$ in charm leptoproduction for $\xi =44$. Plotted are the LO (solid line) 
and NLO (dashed line) FFNS predictions, as well as the ACOT($\chi$) VFNS (dotted curve) results. 
\emph{Right panel:} $x$ dependence of the corresponding predictions for the Callan-Gross
ratio, $R(x,Q^2)=F_L/F_T$, at the same value of $\xi$.}
\end{center}
\end{figure}

The left panel of Fig.~\ref{Fg.2} shows the quantity $F_{2}(x,Q^{2})$ as a function of
$x$ for $\xi =44$. The LO and NLO predictions of the FFNS are given by solid and dashed lines, 
correspondingly.\footnote{Calculating the NLO corrections to the $x$ dependence of quantities 
$F_{2}(x,Q^{2})$ and $R(x,Q^{2})$ presented in Fig.~\ref{Fg.2}, we use the exact results for the 
coefficient functions $c_{k,\mathrm{g}}^{(1,l)}(z,\lambda)$ ($k=2,L$) given in Refs.~\cite{LRSN,RSN}.} 
The ACOT($\chi$) predictions of the VFNS are presented by dotted curve. 
One can see that radiative corrections to the GF mechanism are sizeable, especially for large 
$x\sim 10^{-1}$. At the same time, the difference between the NLO corrections and the charm-initiated
contributions to $F_{2}(x,Q^{2})$ is small: it varies slowly from $15\%$ at low $x\sim 10^{-4}$ 
to $10\%$ at $x\sim 10^{-1}$.

The right panel of Fig.~\ref{Fg.2} shows the $x$ dependence of the Callan-Gross ratio $R(x,Q^2)$ 
for the same value of $\xi$. 
In this case, the NLO corrections are small: they are less than $10\%$ for 
all $x\gtrsim 10^{-4}$. However, the charm-initiated contributions lead to a sizeable decreasing 
of the GF predictions for the  ratio $R(x,Q^2)$. One can see from the right panel of Fig.~\ref{Fg.2} 
that the relative difference between 
the dashed and dotted lines varies from $40\%$ at $x\sim 10^{-4}$ to $70\%$ at $x\sim 10^{-1}$. 
The origin of this effect is straightforward: according to Eq.~(\ref{20}), the QS component does not 
contribute to the longitudinal structure function $F_{L}(x,Q^{2})$.
\begin{figure}
\begin{center}
\begin{tabular}{cc}
\mbox{\epsfig{file=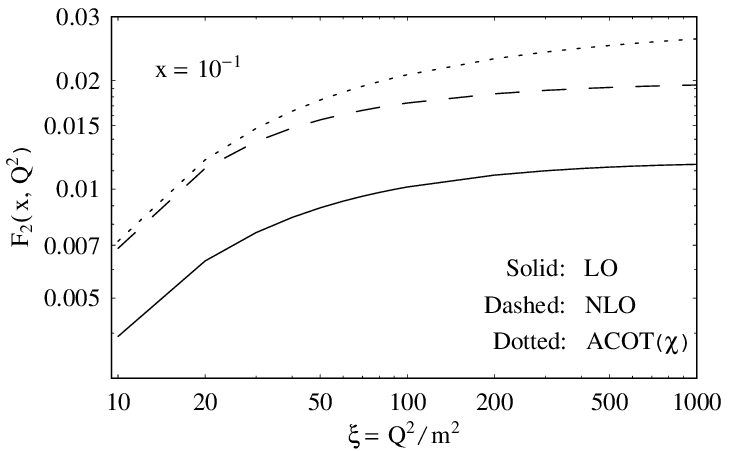,width=250pt}}
& \mbox{\epsfig{file=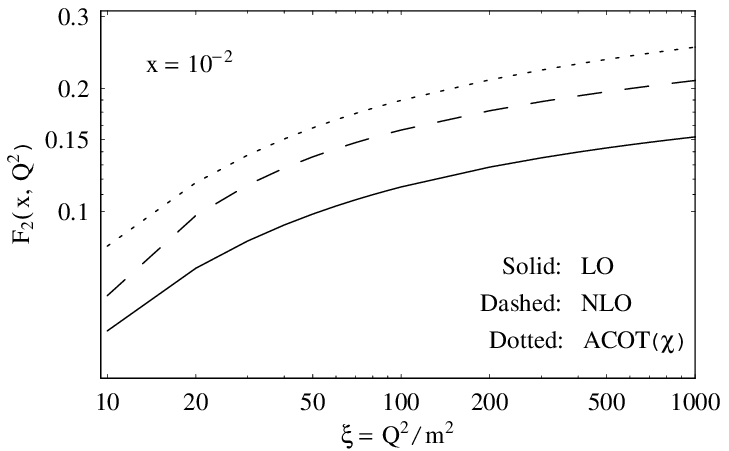,width=250pt}}\\
\mbox{\epsfig{file=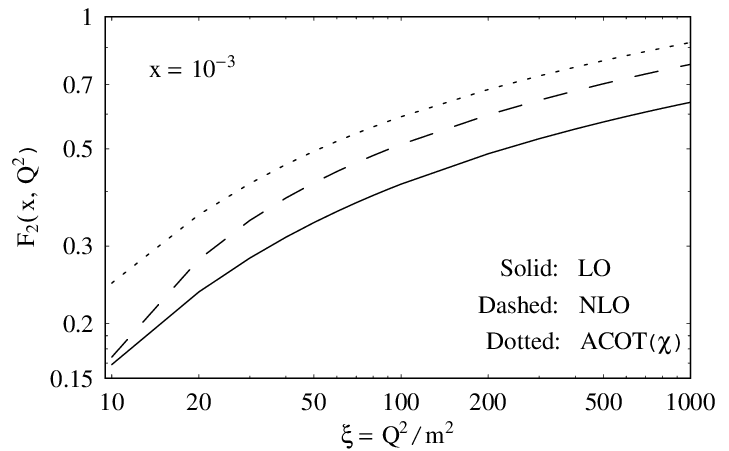,width=250pt}}
& \mbox{\epsfig{file=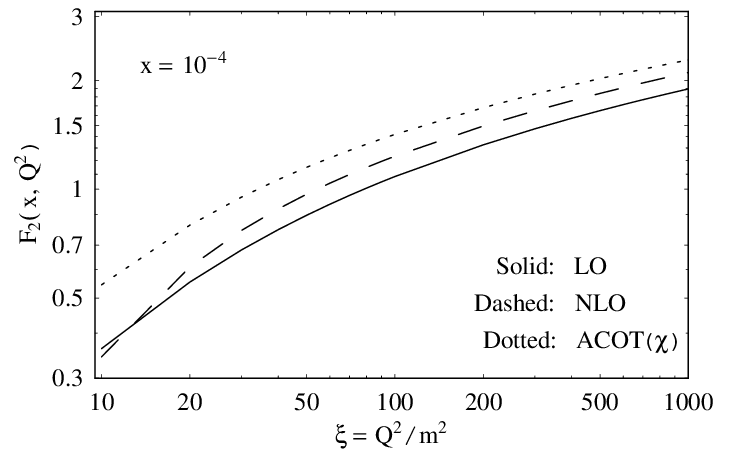,width=250pt}}\\
\end{tabular}
\caption{\label{Fg.3}\small $Q^2$ dependence of the structure function $F_2(x,Q^2)$ in charm 
leptoproduction at $x=10^{-1}$, $10^{-2}$, $10^{-3}$ and $10^{-4}$ for high $Q^2\gg m^2$. 
Plotted are the LO (solid lines) and NLO (dashed lines) FFNS predictions, as well as the 
ACOT($\chi$) VFNS (dotted curves) results.}
\end{center}
\end{figure}

\begin{figure}
\begin{center}
\begin{tabular}{cc}
\mbox{\epsfig{file=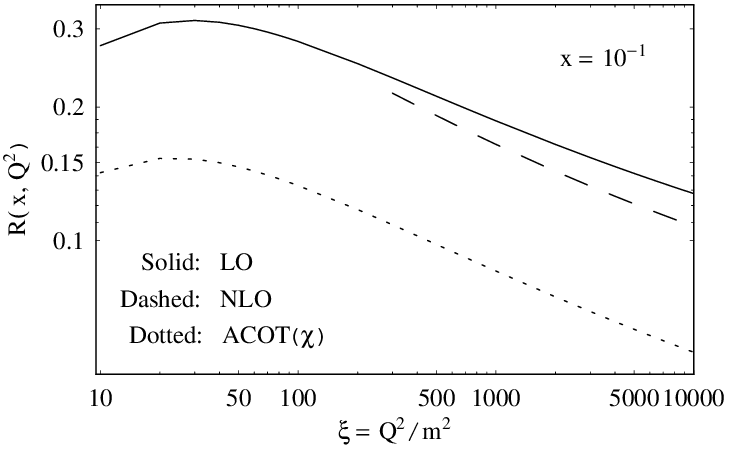,width=250pt}}
& \mbox{\epsfig{file=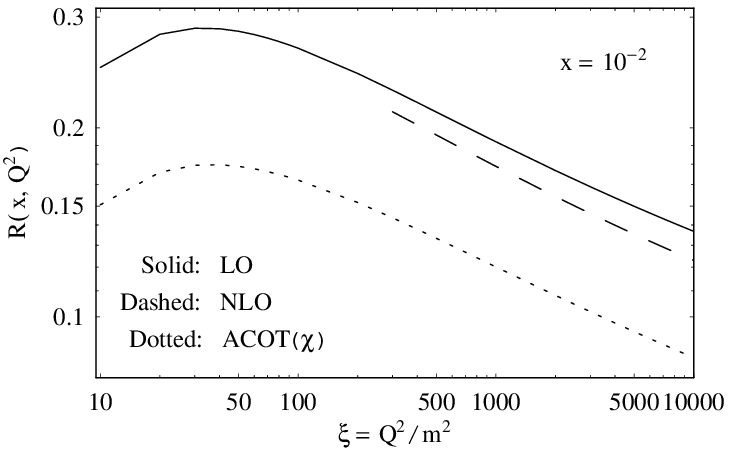,width=250pt}}\\
\mbox{\epsfig{file=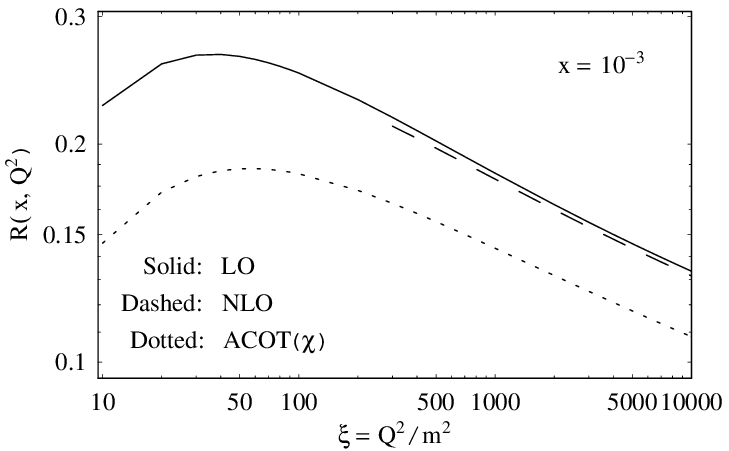,width=250pt}}
& \mbox{\epsfig{file=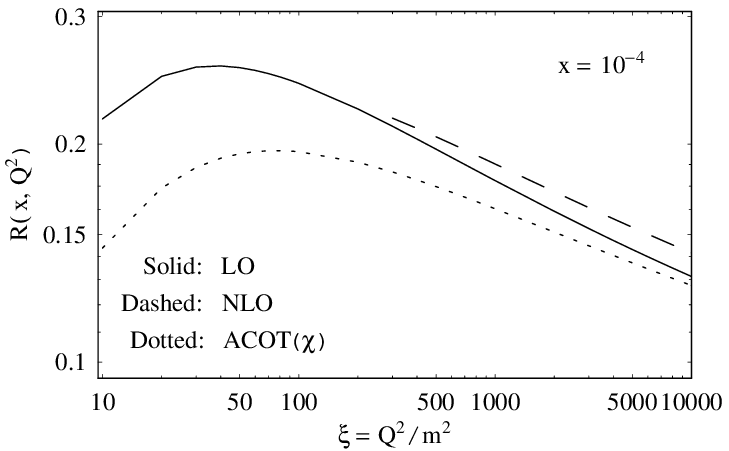,width=250pt}}\\
\end{tabular}
\caption{\label{Fg.4}\small $Q^2$ dependence of the Callan-Gross ratio, $R(x,Q^2)=F_L/F_T$, in 
charm leptoproduction at $x=10^{-1}$, $10^{-2}$, $10^{-3}$ and $10^{-4}$ for high $Q^2\gg m^2$. 
Plotted are the LO (solid lines) and NLO (dashed lines) FFNS predictions, as well as the 
ACOT($\chi$) VFNS (dotted curves) results.}
\end{center}
\end{figure}

The $Q^2$ dependence of the charm-initiated contributions to $F_{2}(x,Q^{2})$ and $R(x,Q^2)$ is 
investigated in Figs.~\ref{Fg.3} and \ref{Fg.4}, respectively. 
Calculating the asymptotic ($Q^2\gg m^2$) NLO predictions, we use the analytic results for the 
coefficient functions $a_{k,\mathrm{g}}^{2,(n,m)}(z)$ ($k=2,L$) presented in Ref.~\cite{BMSMN}.
One can see from Fig.~\ref{Fg.3} that, at $x\sim 10^{-1}$, both the radiative corrections and 
QS contributions to $F_{2}(x,Q^{2})$ are large: they increase the LO GF results by 
approximately a factor of two for all $Q^2$. At the same time, the relative difference between 
the dashed and dotted lines does not exceed $25\%$ for $\xi<10^{3}$.

Considering the corresponding predictions for the ratio $R(x,Q^2)$ presented in Fig.~\ref{Fg.4}, 
we see that, in this case, the NLO and QS contributions are strongly different. The NLO corrections
to $R(x,Q^2)$ are small, less than $15\%$, for $x\sim 10^{-2}$--$10^{-1}$ and $\xi<10^{4}$. On the other 
hand, the corresponding charm-initiated contributions are large: they decrease the GF predictions by 
about $50\%$ practically for all values of $\xi>10$. We conclude that, contrary to the the production 
cross sections, the Callan-Gross ratio $R(x,Q^2)=F_L/F_T$ could be good probe of the charm density in 
the proton at $x\sim 10^{-2}$--$10^{-1}$ and high $Q^2\gg m^2$.

Note that this observation depends weakly on the PDFs we use. We have verified that all the recent CTEQ
versions \cite{CTEQ4,CTEQ5,CTEQ6} of the PDFs lead to a sizeable reduction of the GF 
predictions for the ratio $R(x,Q^2)=F_L/F_T$.

One can also see from Fig.~\ref{Fg.4} that both the radiative and charm-initiated corrections to 
$R(x,Q^2)$ are small, less than $15\%$, for $x\sim 10^{-4}$ and $\xi\sim 10^{3}$--$10^{4}$. 
For this reason, it seems to be difficult to discriminate experimentally between the GF and QS 
contributions at $x\sim 10^{-4}$.

As to the low $x\to 0$ behavior of the Callan-Gross ratio, this problem requires resummation 
of the BFKL terms of the type $\ln (1/x)$ \cite{BFKL1} for both the GF and QS components and 
will be considered in a forthcoming publication.


\section{Conclusion}

We conclude by summarizing our main observations. In the present paper, we compared the structure 
function $F_{2}(x,Q^{2})$ and Callan-Gross ratio $R(x,Q^2)=F_L/F_T$ in charm leptoproduction as 
probes of the charm content of the proton. To estimate the charm-initiated contributions, we used 
the ACOT($\chi$) VFNS \cite{chi}. Our analysis of the radiative and charm-initiated corrections 
indicates that, in a wide kinematic range, both contributions to the structure function 
$F_{2}(x,Q^{2})$ have similar $x$ and $Q^2$ behaviors. For this reason, it is difficult 
to estimate the charm content of the proton using only data on $F_{2}(x,Q^{2})$.

The situation with the Callan-Gross ratio seems to be more optimistic. Our analysis shows that all the 
recent CTEQ versions \cite{CTEQ4,CTEQ5,CTEQ6} of PDFs lead to the VFNS predictions for $R(x,Q^2)$ 
which are about half as large as the corresponding FFNS ones for $x\sim 10^{-2}$--$10^{-1}$ and 
$Q^2\gg m^2$. Taking into account the perturbative stability of the Callan-Gross ratio $R(x,Q^2)=F_L/F_T$ 
within the FFNS \cite{we7}, this fact implies that the charm density in the proton can, in principle, 
be determined from future high-$Q^2$ data on this ratio.

The VFN schemes have been proposed to resum the mass logarithms of the form $\alpha_{s}^{n}\ln^{n}
(Q^{2}/m^{2})$ which dominate the production cross sections at high energies, $Q^2\to \infty$. 
Evidently, were the calculation done to all orders in $\alpha_{s}$, the VFNS and FFNS 
would be exactly equivalent. There is a point of view advocated in Refs.~\cite{ACOT,collins} that,
at high energies, the perturbative series converges better within the VFNS than in the FFNS. There
is also another opinion \cite{BMSN,neerven} that the above logarithms do not vitiate the
convergence of the perturbation expansion so that a resummation is, in principle, not necessary.
Our analysis of the Callan-Gross ratio $R(x,Q^2)=F_L/F_T$ in charm leptoproduction indicates an 
experimental way to resolve this problem. First, contrary to the production cross sections, the 
ratio $R(x,Q^2)=F_L/F_T$ is well defined numerically in FFNS: it is stable both parametrically 
and perturbatively  \cite{we7} in a wide region of $x$ and $Q^2$. Second, it is shown in the present 
paper that the Callan-Gross ratio is very sensitive to resummation of the mass logarithms for 
$x\sim 10^{-2}$--$10^{-1}$ and $Q^2\gg m^2$. Third, nonperturbative contributions 
(like the intrinsic gluon motion in the target) cannot affect both above results at sufficiently 
large $Q^{2}$ where the VFNS is expected to be adequate. Therefore measurements of the Callan-Gross 
ratio in charm leptoproduction would make it possible to clarify the question whether the VFNS 
perturbative series converges better than the FFNS one.

\begin{acknowledgments}
The author thanks S.J. Brodsky for drawing his attention to the problem considered in this
paper.
I am also grateful to B.A.~Kniehl for interesting and useful discussions.

\end{acknowledgments}

\end{document}